\documentstyle[12pt,aaspp4]{article}
\begin{document}

\title{Testing the COBE/IRAS All-Sky Reddening Map Using the Galactic
Globular Clusters}

\author{K. Z. Stanek\altaffilmark{1}}
\affil{Harvard-Smithsonian Center for Astrophysics, 60 Garden St.,
MS20, Cambridge, MA 02138}
\affil{e-mail: kstanek@cfa.harvard.edu}
\altaffiltext{1}{On leave from N.~Copernicus Astronomical Center, 
Bartycka 18, Warszawa 00--716, Poland} 

\begin{abstract}

We live in a dusty Universe, and correcting for the dust extinction
and reddening affects almost all aspects of the optical astronomy.
Recently Schlegel, Finkbeiner \& Davis published an all-sky reddening
map based on the COBE/DIRBE and IRAS/ISSA infrared sky surveys. Their
map is intended to supersede the older Burstein \& Heiles reddening
estimates. In this paper I test this new reddening map by comparing
the reddening values for a sample of 110 $|b|>5\deg$ Galactic globular
clusters selected from compilation of Harris.  I find a good agreement
for globular clusters with galactic latitude $|b|>20\deg$ and fair
overall agreement for globular clusters with $20>|b|>5\deg$, but with
several significant deviations. I discuss four individual clusters
with largest deviations, NGC 6144, Terzan 3, NGC 6355 and IC 1276, in
order to investigate the reasons for these large deviations. It seems
that the new reddening map overestimates the reddening in some large
extinction regions.  However, with its high spatial resolution the new
reddening map can be used to estimate the relative variation of the
reddening on scales $\lesssim 10'$.

\end{abstract}


\section{INTRODUCTION}

We live in a dusty Universe (Hoover 1998, private communication), and
correcting for the dust extinction and reddening affects almost all
aspects of optical astronomy. For us, observing from within the Milky
Way, it is of crucial importance to know how much Galactic dust there
is towards various objects. Burstein \& Hailes (1982; hereafter: BH)
constructed an all-sky reddening map, used extensively by the
astronomical community.\footnote{Their paper was cited 540 times
between 1992 and 1997} Recently, Schlegel, Finkbeiner \& Davis (1998;
hereafter: SFD) published a new all-sky reddening map, based on the
COBE/DIRBE and IRAS/ISSA maps.\footnote{The reddening map and related
files and programs are available using the {\tt WWW} at: {\tt
http://astro.berkeley.edu/davis/dust/}} This map is intended to
supersede the BH map in both the accuracy (16\%) and the spatial
resolution ($6.1'$). Indeed, the potential of the SFD reddening map is
immediately apparent after examining their Fig.7. It is therefore
important to independently test the SFD map to determine the accuracy
of the predicted reddenings.

In this paper I use an electronic catalog of the Galactic globular
clusters compiled by Harris (1996) to test the SFD map.  In Section~2
I compare the values of reddening and discuss the dependence of the
deviations on various properties of the GCs. In Section~3 I discuss
several individual GCs for which the derived deviations are the
largest.

\section{COMPARISON BETWEEN THE REDDENING VALUES}

Harris (1996) surveyed the vast literature on the Galactic globular
clusters (GCs), producing an electronic catalog\footnote{The catalog
is available using the {\tt WWW} at: {\tt
http://www.physics.mcmaster.ca/Globular.html}} of GCs with reasonably
well-known properties, among them the reddening $E(B-V)$.  Many
Galactic globular clusters are extensively studied objects, for which
it is possible to accurately determine their extinction and reddening
using a variety of methods (Burstein \& Heiles 1978; Webbink 1985;
Zinn 1985; Reed, Hesser \& Shawl 1988). However, many GCs in Harris
(1996) compilation have been poorly studied, in which cases the
reddening estimates are rather inaccurate.

The additional advantage of GCs comes from their distribution on the
sky, which, although concentrated towards the Galactic center, has a
sizeable number of GCs scattered across the whole sky. This assures
that there is s good sample of GCs far enough from us that we
``intercept'' most of the interstellar reddening when observing a
cluster. Burstein \& Heiles (1978) used a sample of 49 globular
clusters with $|b|>10\deg$ to test their method of the reddening
determination.

There are 147 GCs in the catalog of Harris (1996), out of which I
excluded two which had no distance information.  For the remaining 145
clusters I used the program ``dust\_getval'' included with the SFD
reddening map to obtain the value of reddening predicted by the SFD,
which I hereafter call $E(B-V)_{SFD}$. This was done by obtaining 49
values of $E(B-V)_{SFD}$ on a uniform $7\times7$ grid with $2.5'$
spacing centered on the cluster ($15'\times15'$ part of the sky),
which allowed me to obtain an average value of $E(B-V)_{SFD}$ as well
as its standard deviation $\sigma_{E(B-V)}$.  At this point I excluded
35 GCs with 'no\_list' flag, which were mostly clusters with the
galactic latitude $|b|<5\deg$, but also two GCs which fell into the
region of the sky not scanned by IRAS.  This leaves me with 110 GCs
with which I can test the SFD map. I denote the reddening obtained
using the catalog of Harris by $E(B-V)_H$.

\begin{figure}[t]
\plotfiddle{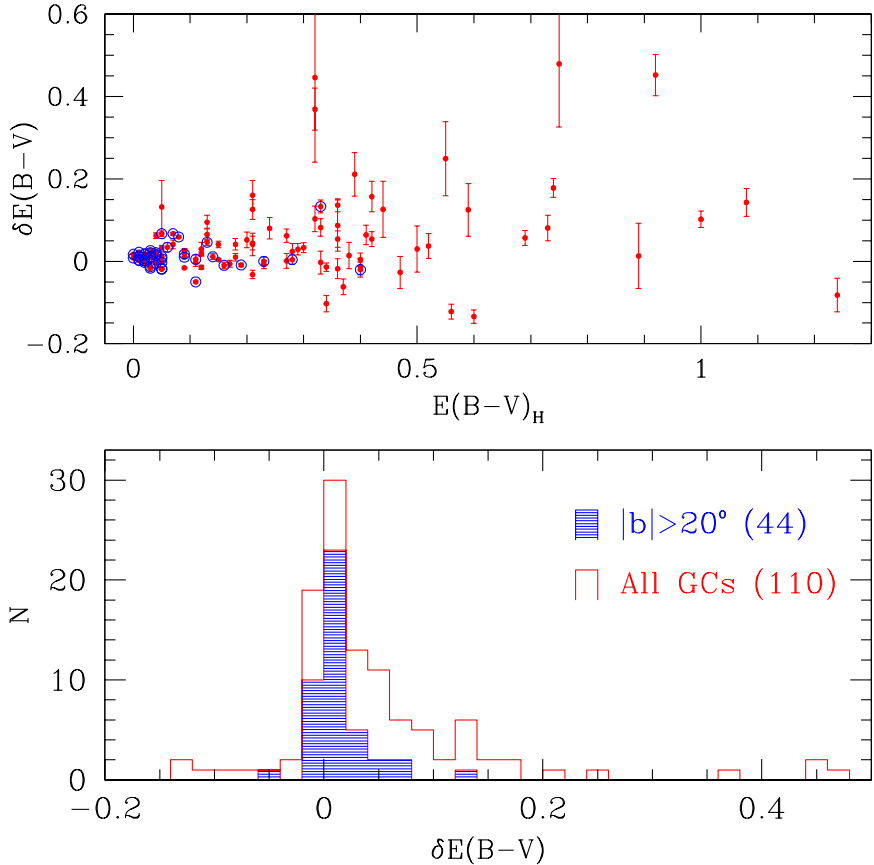}{9.6cm}{0}{60}{60}{-190}{-105}
\caption{Deviations between the values of the reddening $E(B-V)_H$
obtained using the electronic catalog of Harris (1996) and the
reddening $E(B-V)_{SFD}$ obtained using the all-sky map of Schlegel,
Finkbeiner \& Davis~(1998). In the upper panel I plot the deviations
$\delta E(B-V)\equiv E(B-V)_{SFD}-E(B-V)_H$ as the function of
$E(B-V)_H$.  Larger symbols denote clusters with $|b|>20\deg$. In the
lower panel I plot the histogram of the $\delta E(B-V)$ values for
all 110 Galactic globular clusters and for the subset of 44 GCs with
$|b|>20\deg$ (shaded).}
\label{fig:sfd}
\end{figure}

In the upper panel of Figure~\ref{fig:sfd} I plot the values of the deviations
between the SFD values and the Harris' values $\delta E(B-V)\equiv
E(B-V)_{SFD}-E(B-V)_H$ as the function of $E(B-V)_H$, with 
$\sigma_{E(B-V)}$ as errorbars.  In the lower panel of Figure~\ref{fig:sfd} I show
the histogram of the deviations $\delta E(B-V)$ for all 110 Galactic
globular clusters and for the subset of 44 GCs with $|b|>20\deg$
(shaded). The reddening $E(B-V)_H$ extends from 0.0 to 1.24 and the
deviations cover $-0.13<\delta E(B-V)<0.48$.

The distribution of $\delta E(B-V)$ is skewed towards the positive
values, which might be to some extent expected. The SFD map gives, by
its construction, a total dust emission (converted to reddening) along
given line of sight, so it could be expected that the GCs, located at
some finite distance from us, are subject to only some fraction of the
total reddening. The subset of GCs with $|b|>20\deg$ shows a much
smaller values of $\delta E(B-V)$, with only one cluster with $\delta
E(B-V)>0.1\;$mag.  However, the $|b|<20\deg$ sample shows a
significant number of large deviations $\delta E(B-V)>0.1\;$mag,
including four clusters with $\delta E(B-V)>0.3\;$mag. This can result
from number of effects:

\begin{enumerate}

\item{} Inaccurately determined $E(B-V)_{SFD}$ for the $|b|<20\deg$
region of the sky (or part of it containing large number of GCs,
i.e. region close to the Galactic center), which on average is subject
to significantly more dust reddening than the high galactic latitude
regions (Figure~\ref{fig:sfd}, upper panel). As noted by SFD (their Fig.6) when
calibrating the COBE/IRAS map, their highest reddening values appear
to be overestimated. This could be caused by the ratio $R_V\equiv
A_V/E(B-V)$ being significantly different from the value of 3.1,
assumed to hold universally by SFD, in some high extinction regions.

\item{} Poorly determined $E(B-V)_H$ for some of the $|b|<20\deg$
clusters.

\item{} The finite distance to the globular clusters, which for small
values of the galactic latitude $b$ could translate to some Galactic
dust behind the clusters.

\end{enumerate}

It is possible that all of the above contribute to the large
deviations $\delta E(B-V)>0.1\;$mag found. In the next Section I will
discuss in some detail the four clusters with the largest $\delta
E(B-V)$ deviations trying to determine which effect might be the
dominant one.

\section{INVESTIGATING THE DEVIATIONS}

\begin{figure}[t]
\plotfiddle{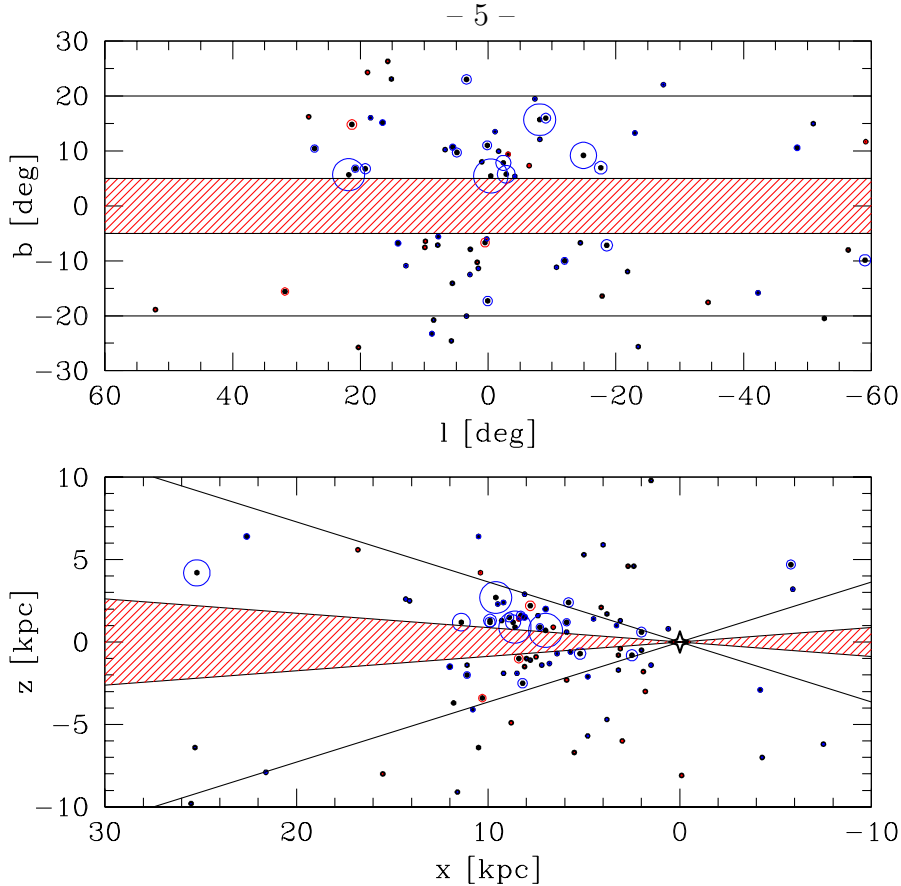}{9.6cm}{0}{60}{60}{-190}{-105}
\caption{In the upper panel I show $120\times60\;\deg$ part of the
sky including the Galactic center, with the globular clusters plotted
as small dots. For each cluster I also plot a circle with radius
proportional to its $|\delta E(B-V)|$, with the largest circle
corresponding to $\delta E(B-V)=0.48\;$mag. In the lower panel I plot
the GCs in the x-z plane, perpendicular to the plane of the Galaxy,
with the Sun at $(x,z)=(0,0)$. The symbols are the same as in the
upper panel. The shaded regions correspond to the $|b|<5\;deg$ part of
the Galaxy, from which the globular clusters were removed.}
\label{fig:xz}
\end{figure}

In the upper panel of Figure~\ref{fig:xz} I show $120\times60\;\deg$ part of the
sky centered on the Galactic center, with the globular clusters
plotted as small dots. For each cluster I also plot a circle with
radius proportional to its $|\delta E(B-V)|$, with the largest circle
corresponding to $\delta E(B-V)=0.48\;$mag. The shaded region
correspond to the $|b|<5\;deg$ part of the Galaxy, from which the
globular clusters were removed. All four GCs with $\delta
E(B-V)>0.3\;$mag are located in the northern Galactic hemisphere and
are relatively close to each other on the sky.

\begin{planotable}{lccrrrrr}
\tablewidth{42pc}
\tablecaption{Globular Clusters With Largest Deviations $\delta E(B-V)$}
\tablehead{ \colhead{Name} & \colhead{$E(B-V)_H$} & \colhead{$E(B-V)_{SFD}$} 
& \colhead{$l$} & \colhead{$b$} & \colhead{$R_{\odot}$}   
& \colhead{$x$} & \colhead{$z$} \\ 
\colhead{ } & \colhead{$[mag]$} &  \colhead{$[mag]$} &
\colhead{$[\deg]$} & \colhead{$[\deg]$}  & \colhead{$[kpc]$} 
& \colhead{$[kpc]$} & \colhead{$[kpc]$} }
\startdata
NGC 6144  & 0.32 & $0.77\pm0.21$ & 351.9286 & 15.6990 & 10.1 &  9.6 & 2.7 \nl
Terzan 3  & 0.32 & $0.69\pm0.05$ & 345.0765 &  9.1868 & 26.4 & 25.2 & 4.2 \nl 
NGC 6355  & 0.75 & $1.23\pm0.15$ & 359.5849 &  5.4277 &  7.1 &  7.0 & 0.7 \nl
IC 1276   & 0.92 & $1.37\pm0.05$ &  21.8321 &  5.6685 &  9.3 &  8.6 & 0.9  
\enddata 
\label{table:gc}
\end{planotable}

In the lower panel of Figure~\ref{fig:xz} I plot the GCs in the x-z
plane, perpendicular to the plane of the Galaxy, with the Sun at
$(x,z)= (0,0)$. The symbols are the same as in the upper panel of
Figure~\ref{fig:xz}.  The clusters are concentrated towards the
Galactic center, and so are the GCs with the largest values of
$|\delta E(B-V)|$, most of them positive. This could indicate that the
cause of the large deviations found is the presence of significant
amounts of dust beyond the Galactic center, causing a real difference
between the reddening values measured towards some of the GCs and the
total reddening along these lines of sight.  However, at $\sim 8\;kpc$
from the Sun the galactic latitude of $|b|=5\;deg$ corresponds to the
distance from the Galactic plane of $|z|=700\;pc$, which is
significantly larger than expected for significant amounts of dust to
be present (e.g.~M\'endez \& van Altena 1998). However, as argued by
SFD in their Section 7.3, it might be plausible that such high-$z$
material exists.

I select four GCs with the largest deviations $\delta E(B-V)>
0.3\;$mag (all of them positive). Their properties, selected from
Harris (1996), are in Table~1.  For each cluster I give its name, both
values of reddening $E(B-V)_H$ and $E(B-V)_{SFD}$ (along with its
standard deviation), galactic longitude and latitude $l,b$, distance
from the Sun $R_{\odot}$ and the distance components $x,z$ in a
Sun-centered coordinate system, where $x$ points toward the Galactic
center and $z$ toward the North Galactic Pole. I will now discuss each
cluster in some detail trying to determine the possible cause for the
large $\delta E(B-V)$.

\paragraph{NGC 6144} The value of $E(B-V)_H=0.32$ for this cluster comes
originally from Bica \& Pastoriza (1983), who used an average of their
value based on the integrated photometry and value of $E(B-V)=0.4$ of
Alcaino (1980), based on color-magnitude data. Alcaino (1980) noted
that the extinction for this cluster increases towards the NE
direction. Indeed, using the SFD map for a $15'\times15'$ grid
discussed earlier in the paper I find gradient of $E(B-V)_{SFD}$ from
0.48 to 1.11. Given this large range of $E(B-V)_{SFD}$, large
deviation $\delta E(B-V)\approx 0.45$ for this cluster could be
spurious. As noted by Webbink (1985), many clusters with
$E(B-V)\gtrsim 0.2$ suffer nonuniform extinction, which is confirmed
by large values of $\sigma_{E(B-V)}$ for these clusters (Figure~\ref{fig:sfd}). As it
turns out both NGC 6144 and a neighboring M4 globular cluster are
located on the sky close to the Ophiuchus molecular cloud complex (de
Geus, Bronfman \& Thaddeus 1990). Huterer, Sasselov \& Schechter
(1995) analyzed the consistency of two distance estimates for M4 and
found that it requires a non-standard ratio of total to selective
extinction $R_V=4.0\pm0.2$.

\paragraph{Terzan 3} This is a poorly studied globular cluster
with no color-magnitude data, for which the $E(B-V)_H$ was estimated
by Webbink (1985) using a modified ``cosecant law'', an educated guess
at best.  It seems therefore justified to replace $E(B-V)_H$ by the
value of $E(B-V)_{SFD}=0.69\pm0.05$, which has the effect of reducing
its $V$-band distance modulus by $\sim 1.2\;$mag, and therefore
reducing its distance $R_{\odot}$ from $25.2\;kpc$ to $14.5\;kpc$.

\paragraph{NGC 6355} The value of $E(B-V)_H=0.73$ was adopted
by Webbink (1985), based on integrated photometry of the cluster,
which as discussed by Webbink is a secondary method of medium
accuracy. As for NGC 6144, the SFD map gives a large gradient of
$E(B-V)$ from 0.96 to 1.50 on the $15'\times15'$ grid centered on the
cluster.

\paragraph{IC 1276} The value of $E(B-V)_H=0.92$ was adopted
by Webbink (1985), based on integrated photometry of the cluster.  The
SFD map gives only a modest range of $E(B-V)$ from 1.26 to 1.46.

\section{DISCUSSION}

\begin{figure}[t]
\plotfiddle{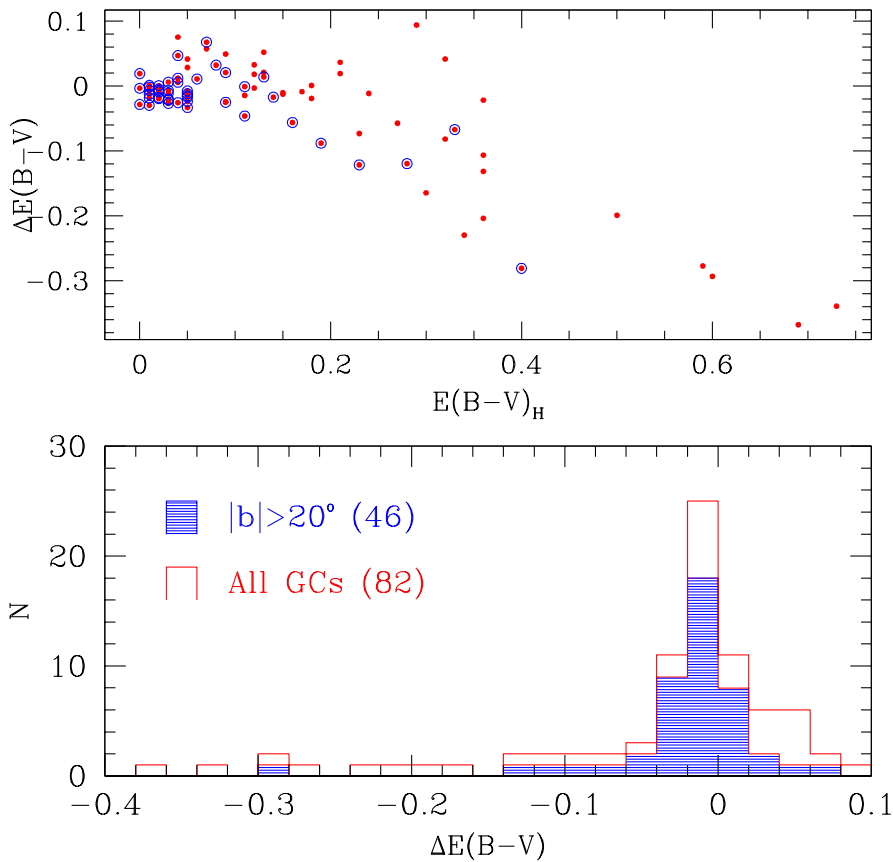}{9.6cm}{0}{60}{60}{-190}{-105}
\caption{Deviations between the values of the reddening $E(B-V)_H$
obtained using the electronic catalog of Harris (1996) and the
reddening $E(B-V)_{BH}$ obtained using the reddening map of Burstein
\& Hailes~(1982). In the upper panel I plot the deviations $\Delta
E(B-V)\equiv E(B-V)_{BH}-E(B-V)_H$ as the function of $E(B-V)_H$.
Larger symbols denote clusters with $|b|>20\deg$. In the lower panel I
plot the histogram of the $\Delta E(B-V)$ values for 82 Galactic
globular clusters with $|b|>10\deg$ and for the subset of 46 GCs with
$|b|>20\deg$ (shaded).}
\label{fig:bh}
\end{figure}

It might be also interesting to compare the deviations between the
values of $E(B-V)_H$ and those predicted by the BH map, which I denote
$E(B-V)_{BH}$. In Figure~\ref{fig:bh}, analogous to
Figure~\ref{fig:sfd}, I plot the deviation $\Delta E(B-V)\equiv
E(B-V)_{BH}-E(B-V)_H$. Since the BH map excludes objects with
$|b|<10\deg$, there is fewer (82) GCs now, with smaller range of
$E(B-V)_H$. There is a clear trend between $E(B-V)_H$ and $\Delta
E(B-V)$, indicating that the BH map underestimates the value of
reddening, especially in high reddening regions. As noticed by SFD,
even in high galactic latitude regions there is a systematic offset of
$\sim 0.02\;$mag between the SFD and the BH maps, with the BH map
predicting lower values of the reddening.  Comparing
Figures~\ref{fig:sfd} and \ref{fig:bh}, it is clear that overall the
SFD map does a much better job in predicting the reddening values of
the Galactic globular clusters.

To summarize, the SFD map predicts well the reddening for most of the
globular clusters in the Harris (1996) sample, especially for the
$|b|>20\deg$ clusters. It is however possible that it overestimates
the reddening in some large extinction regions, such as in the case of
NGC 6144, NGC 6355 and IC 1276 discussed above, although it might be
that the $E(B-V)$ estimates based on the integrated properties of
globular clusters are at fault here (see discussion in Burstein \&
Hailes 1978).  In any case, the SFD map is a very valuable tool in
predicting large reddening gradients on scales of $\lesssim
10'$. Further tests of the SFD map would be most useful, using for
example field RR Lyrae stars (Burstein \& Hailes 1978), such as the
sample of Layden (1998), but with reddenings not based on the BH map.

\acknowledgments{I was supported by the Harvard-Smithsonian Center for
Astrophysics Fellowship. I am happy to acknowledge helpful discussions
with Dimitar Sasselov and Pat Thaddeus.  David Schlegel provided me
with helpful comments on the manuscript, and also suggested comparison
with the BH map. This research has made use of NASA's Astrophysics
Data System Abstract Service.}

\end{document}